\documentclass[aps,prb,twocolumn,groupedaddress]{revtex4}
\usepackage{epsfig}

\usepackage{amsmath}

\begin{document}
\title{Interplay between the pseudogap and superconductivity in underdoped HgBa$_2$CuO$_{4+\delta}$ single crystals}
\author{Y. Gallais$^{1,2}$, A. Sacuto$^{1,2}$, T. P. Devereaux$^{3}$ and D. Colson$^{4}$}
\address{$^{1}$Laboratoire de Physique du Solide (UPR 5 CNRS) ESPCI, 10 rue
Vauquelin
75231 Paris, France\\
$^{2}$Mat\'eriaux et Ph\'enom\`enes Quantiques (FDR 2437 CNRS),
Universit\'e Paris 7, 2 place Jussieu 75251 Paris.\\
$^{3}$Departement of Physics, University of Waterloo, Waterloo, Ontario, Canada N2L 3G1 \\
$^{4}$Service de Physique de l'Etat Condens\'{e}, CEA-Saclay,
91191 Gif-sur-Yvette, France}
\date{\today}

\begin{abstract}
We report a doping dependant Electronic Raman Scattering (ERS)
study of HgBa$_2$CuO$_{4+\delta}$ (Hg-1201) single crystals. We
investigate the dynamics of the antinodal and nodal
quasiparticles. We show that the dynamical response of the
antinodal quasiparticles is strongly reduced towards the
underdoped regime in both the normal and superconducting states.
When probing the nodal quasiparticles, we are able to distinguish
between the energy scale of the pseudogap and that of the
superconducting gap. A simple model relating the suppression of
the dynamical response of the antinodal quasiparticles to
fluctuations related to a competing phase is proposed.
\end{abstract}
\maketitle Among the unsolved problems of the high temperature
superconductivity field, the question of the nature of the
pseudogap phase is certainly the most highly debated. In
particular its relationship with superconductivity remains an open
question. One class of model attributes the pseudogap phase as a
precursor phase of superconductivity in which Cooper pairs form at
a temperature $T^*$ but only acquire phase coherence at a lower
temperature $T_c$ where they form an uniform $d$-wave BCS
condensate.\cite{Emery-Kivelson} An important consequence of all
these theories is that the pseudogap and the superconducting gap
are intimately connected: they can be both identified to a pairing
energy. This approach is supported by ARPES data which show that
the pseudogap and the superconducting gap open in the same region
of the Fermi surface,\cite{Ding} i.e. close to the ($\pi$,0) and
related points. Another class of models invokes various phases
which are not directly related to superconductivity but
 rather compete
 with it. Among the proposed phases
 we find a precursor SDW phase,\cite{Pines} a $d$-density wave phase \cite{Chakra} or
 an orbital current phase.\cite{Varma} Most of these theories, but
 not all, predict the presence of a quantum critical point somewhere near
 the optimally doped regime.\cite{Tallon,Sachdev} Specific heat data, doping evolution of the superfluid
 density and impurity induced $T_c$ suppression, among others, have been interpreted as
 evidences for this scenario.\cite{Tallon}

In order to fully understand the nature of the pseudogap, a
two-particle response function able to probe quasiparticle dynamics on different
regions of the Brillouin zone would be useful. In this report we
show that Electronic Raman Scattering (ERS) is such a probe and
report a doping dependant
study
of the interplay between the pseudogap and superconductivity in
HgBa$_2$CuO$_{4+\delta}$ (Hg-1201) single crystals.
Being a two-particle probe, ERS is able to access the
charge dynamics in both the normal and the superconducting
states.\cite{Naeini,Chen,Nemetschek,Blumberg} Moreover, through
the use of particular sets of incident and scattered
polarizations, it is able to probe different regions of the Fermi
surface, i.e. the nodal (along the (0,0)-($\pi,\pi$) direction)
and antinodal (along the (0,0)-$(\pi,0)$ direction)
quasiparticles. With this unique ability, ERS is thus expected to
give important information on the pseudogap nature which are not
accessible via one-particle probes such as ARPES.

With only one CuO$_2$ plane per unit cell and a pure tetragonal
symmetry, Hg-1201 is a very attractive compound to study the
intrinsic physics of the CuO$_2$ plane in the underdoped regime.
In particular underdoped Hg-1201 is closer to its fully
stoichiometric phase ($\delta$=0) than optimally and over-doped
Hg-1201 and should therefore be structurally more ordered. In this
study, we show that the pseudogap in Hg-1201 manifests itself as a
suppression of the superconducting Raman response along the
antinodal directions. Comparison with available data on other
compounds shows that this suppression is generic to the cuprates
and starts slightly above optimal doping. When probing the nodal
directions, we observe a partial loss of spectral weight in the
response which starts around 700~cm$^{-1}$ in the normal state,
while below $T_c$, the superconducting gap opens at about
200~cm$^{-1}$ thus indicating a non-superconducting origin of the
pseudogap. We propose a simple model which relates the suppression
of the antinodal dynamical response to two-particle vertex
suppression arising from fluctuations of a competing phase.

The Hg-1201 single crystals studied have been successfully grown
by the flux method. The detailed procedure for crystal growth will
be described elsewhere.\cite{Colson} The ERS measurements have
been performed on three as-grown single crystals with different
dopings. The single crystals studied here were carefully selected from several batches
for their sharp transitions widths (less than
5~K). Their magnetically measured transition temperatures, T$_c$,
are 95~K, 78~K and 63~K. We will refer to these
crystals as Hg95K, Hg78K and Hg63K respectively. The first sample
is very close to optimal doping while the latter two are
underdoped. The as-grown surfaces of the single crystals were
mechanically polished to suppress small extrinsic impurity phases
which are known to exist at the surface of mercurate compounds.\cite{Sacuto-imp} The spectra were obtained using the 514.5~nm
(2.4~eV) excitation line of an Ar$^+$-Kr$^+$ laser. The scattered
light was analysed using a triple grating spectrometer (JY-T64000)
equipped with a nitrogen cooled CCD detector. In this study we
focus on B$_{1g}$ and B$_{2g}$ symmetries which can be
individually selected using specific configurations of incident
and scattered electric fields with respect to the cristallographic
axis. In the cuprates, the B$_{1g}$ (x'y' scattering geometry) symmetry is
sensitive to regions along the antinodal directions while the
B$_{2g}$ (xy scattering geometry) symmetry probes mostly along the nodal
directions. The spectra presented here have been corrected for the
spectral response of the spectrometer and for the Bose-Einstein
factor. They are thus proportional to the imaginary part of the
Raman response function $\chi''$. All the referred temperature
have been corrected for the estimated laser heating.

\begin{figure}
\centering \epsfig{figure=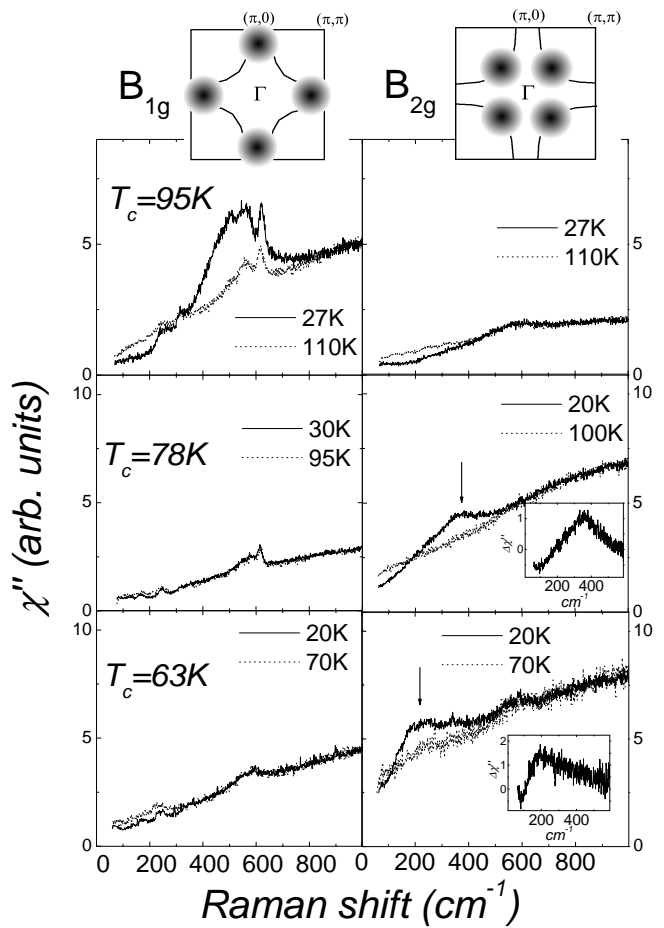, width=0.99\linewidth,
clip=} \caption{Raman spectra of Hg-1201 at various doping levels
in B$_{1g}$ and B$_{2g}$ symmetries. Spectra are shown for
temperatures below and slightly above T$_c$. The regions probed in
B$_{1g}$ and B$_{2g}$ symmetries are displayed on a schematic
Fermi surface for hole-doped cuprates in the upper side. Insets
show the difference between the superconducting and normal states
spectra, $\Delta\chi''$, for Hg78K and Hg63K in the B$_{2g}$
symmetry.} \label{Spectres}
\end{figure}

In FIG. \ref{Spectres} we show the spectra in B$_{1g}$ and
B$_{2g}$ symmetries as a function of doping both above and below
$T_c$. The spectra for the optimally doped sample Hg95K are
consistent with a d-wave superconducting gap and compare well with
spectra of other optimally doped cuprates (for more details, see
ref. 14). The B$_{1g}$ spectrum shows a redistribution
into a 2$\Delta_0$ pair breaking peak upon entering the
superconducting state (2$\Delta_0\sim520~cm^{-1}$). On the other
hand, the spectrum in the B$_{2g}$ symmetry displays a loss of
spectral weight at low frequency but does not show any
superconductivity induced peak. We note that for a $d$-wave gap,
the B$_{2g}$ pair breaking peak intensity is expected to be much
weaker than the B$_{1g}$ one and the peak should be located at a
lower frequency, i.e. below $2\Delta_0$.
As we lower the doping, the response in the B$_{1g}$ symmetry
displays a dramatic suppression of the pair breaking peak
intensity and the normal and superconducting state spectra are
virtually identical. By contrast, the B$_{2g}$ spectra show clear
superconductivity induced peaks in both underdoped samples. The
peaks are located around 360~cm$^{-1}$ and 200~cm$^{-1}$ for Hg78K
and Hg63K respectively and are associated to the opening of the
superconducting gap. It is well known that the B$_{2g}$ peak
height is largely controlled by disorder.\cite{Dev95,Nemetschek}
The absence of a B$_{2g}$ peak in Hg95K is therefore most likely
related to the fact that optimally doped Hg95K is structurally
more disordered than the underdoped Hg78K and Hg63K. This is
further confirmed by NMR measurements on Hg-1201 which show that
$^{17}$O NMR linewidth increases when going from under to
overdoped samples.\cite{Bobroff97} In fact it is remarkable that
despite the complete disappearance of any coherent B$_{1g}$
superconducting response, the B$_{2g}$ response displays a clear
coherent superconducting response for both underdoped samples.
\begin{figure}
\centering \epsfig{figure=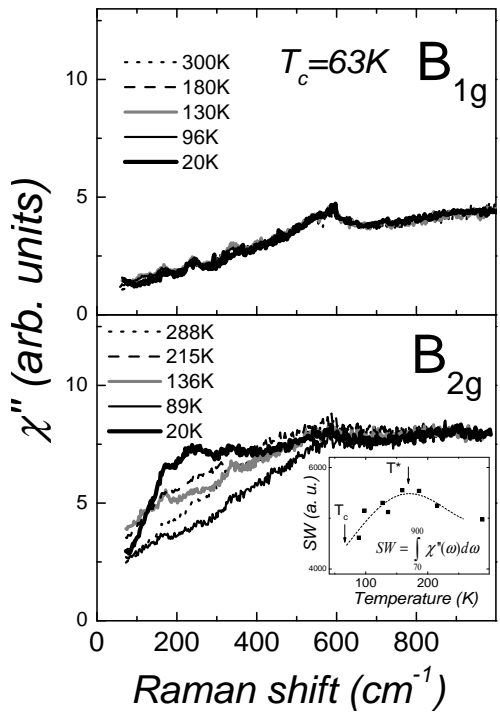, width=0.75\linewidth,clip=}
\caption{Temperature dependance of the B$_{1g}$ (upper panel) and
B$_{2g}$ (lower panel) spectra for Hg63K between room temperature
and 20~K. The inset shows the integrated spectral weight (SW) as a
function of temperature in the B$_{2g}$ channel. It is defined as
SW=$\int^{900}_{70}\chi''(\omega)d\omega$.} \label{B2g(T)}
\end{figure}

In FIG. \ref{B2g(T)} the B$_{1g}$ and B$_{2g}$ responses of Hg63K
are shown as a function of temperature. The first striking result
is that the B$_{1g}$ spectra displays hardly any temperature
dependance when cooling from 300~K down to 20~K. This suggests a
highly incoherent response in this symmetry at least up to
1000~cm$^{-1}$. By contrast the B$_{2g}$ response displays a
complex temperature dependence when cooling from 288~K to 20~K.
Between 288~K and 215~K the response shows conventional metallic
behavior, i.e. an overall increase of the spectral weight at low
frequency. Between 215~K and 136~K, the response develops a
partial suppression which starts below 700~cm$^{-1}$ and further
deepens with cooling. The suppression is only partial because the
215~K and 136~K spectra cross again at about 130~cm$^{-1}$
consistently with a metallic-like behavior at very low frequency.
We identify this suppression as the opening of a anisotropic
pseudogap which leaves parts of the Fermi surface, along the nodal
directions, essentially unaffected. A similar conclusion was
reached by Nemetschek et al. from their analysis of the B$_{2g}$
response in underdoped Y-123 ortho II.\cite{Nemetschek} In the
inset of FIG. \ref{B2g(T)} we plot the evolution of the integrated
spectral weight (SW) as a function of temperature. The integrated
SW increases until a characteristic temperature T* (T*$\sim$170~K)
where it starts to decrease until T$_c$ is reached. The deduced T*
is very close to the one reported by Itoh et al. based on their
NMR data ($^{63}$1/T$_{1}T$) on similarly doped Hg-1201 samples.
\cite{Itoh} The appearance of a pair breaking below T$_c$ in the
same symmetry allows for a direct comparison between the pseudogap
and the superconducting gap. It shows very different associated
energies: the pair breaking peak is located at 200~cm$^{-1}$ while
the pseudogap opens around 700~cm$^{-1}$ thus advocating for a
non-superconducting origin of the pseudogap.

\begin{figure} \centering
\epsfig{figure=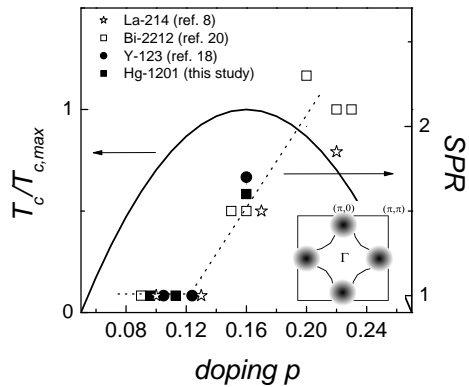,width=0.7\linewidth,clip=}
\caption{Summary of the B$_{1g}$ Superconducting Peak Ratio (SPR)
observed by ERS in various cuprates as a function of doping $p$
.\cite{Naeini,Venturini2001,GallaisY123} The doping was simply
deduced from the ratio $T_c/T_{c,max}$ assuming p$_{opt}$=0.16 as
in La-214.\cite{Presland}} \label{UD-graph}
\end{figure}

We now discuss the striking evolution of the B$_{1g}$ pair
breaking peak intensity with doping. The rapid disappearance of
the B$_{1g}$ peak on the underdoped side of the phase diagram has
also been reported for La-214, Bi-2212 and Y-123. Like the partial
suppression observed in B$_{2g}$ symmetry, it was attributed to
the presence of a pseudogap along the antinodal directions.\cite{Chen}
 More precisely, a destruction of the Fermi surface at
the antinodal points was invoked.
It is important however
to stress that recent ARPES data show clear quasiparticle peaks at
these points in the superconducting state of underdoped
Bi(Pb)-2212, in apparent contradiction with Raman data in the
superconducting state of underdoped cuprates.\cite{Dresde} To
quantify the suppression of the pair breaking peak intensity as we
lower the doping, we have calculated the Superconducting Peak
Ratio (SPR) defined as
SPR=$\chi''_S(\omega=2\Delta_0)/\chi''_N(\omega=2\Delta_0)$ where
$\chi''_S$ and $\chi''_N$ are the Raman responses in the
superconducting and normal states respectively. In FIG.
\ref{UD-graph} this ratio is plotted as a function of doping for
various cuprates. The suppression is clearly a generic property of
the hole-doped cuprates.
In fact, the evolution of the SPR as a function of doping is
strikingly similar to the evolution of the condensation energy and
the c-axis superfluid density in various cuprates.\cite{Tallon,Pana}
FIG. \ref{UD-graph} suggests that the
pseudogap results from interactions which progressively destroy
the superconducting response near the antinodal directions and
leave the nodal directions unaffected at low energy. A similar
picture has emerged recently from STM experiments where
coexistence between nodal superconductivity and charge ordering
involving antinodal quasiparticles was found.\cite{Davis}

In the following we consider a simple phenomenological model which
can account for the absence of any coherent superconducting
response in the B$_{1g}$ channel of underdoped cuprates. The Raman
response is a two particle probe:
\begin{multline}
\chi''_{\gamma\Gamma}(\Omega)=\frac{2}{N}\sum_{\mathbf{k}}\int
\frac{d\omega}{\pi}[f(\omega)-f(\omega+\Omega)]
\\
Tr\{\gamma_{\mathbf{k}}\hat{\tau}_3\hat{G}''(\mathbf{k},\omega)\hat{\Gamma}_{\mathbf{k}}(\omega,\omega+\Omega)
\hat{G}''(\mathbf{k},\omega+\Omega)\} \label{response}
\end{multline}
where $\hat{G}''$ is the full imaginary part of the Green's
function in the superconducting state. $\gamma$, $\hat{\Gamma}$
are the bare, renormalized Raman vertices, respectively, and
$\hat{\tau}_i$ are the Pauli matrices.
If vertex corrections are ignored
($\hat{\Gamma}_{\mathbf{k}}$=$\gamma_{\mathbf{k}}\hat{\tau}_3$)
the Raman response can be calculated from the knowledge of single
particle properties such as the self-energy
$\hat{\Sigma}_{\mathbf{k}}$ and the superconducting gap
$\Delta_{\mathbf{k}}$. While expression (\ref{response}) has been
used extensively to fit Raman spectra in optimally doped cuprates,
the situation in the underdoped cuprates does not fit this simple
picture since no pair breaking peak is observed in the B$_{1g}$
symmetry and
the single particle properties deduced from ARPES are not strongly
doping dependant.\cite{Dresde}
It appears therefore that the rapid decrease of the
superconducting response in the B$_{1g}$ symmetry in the
underdoped regime cannot be accounted by single particle
properties alone.

Thus, in the following, we consider the effect of two-particle
vertex renormalizations to $\hat{\Gamma}_{\mathbf{k}}$. The
renormalized vertex $\hat{\Gamma}_{\mathbf{k}}$ may be calculated
in a number of contexts corresponding to various competing phases.
Due to the topology of the Fermi surface in hole doped cuprates,
it can be shown that, while the B$_{2g}$ vertex amplitude is only
mildly affected by charge suppressing fluctuations centered around
($\pi,\pi$), the B$_{1g}$ vertex is strongly suppressed.
\cite{Dev99} These fluctuations may be related to
antiferromagnetic (AF) order but various competing phases
like the $d$-density wave phase,\cite{Chakra} are expected to
induce a similar suppression.\cite{Dev99}
\begin{figure}[t]
\centering \epsfig{figure=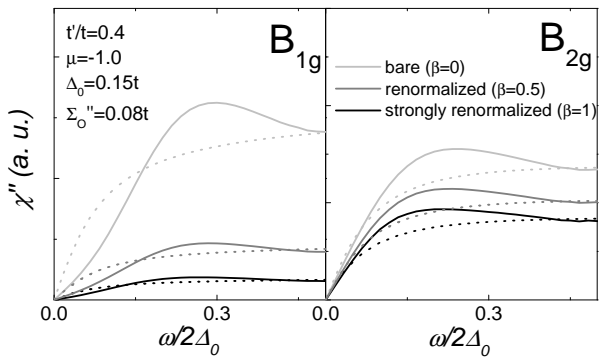, width=0.91\linewidth, clip=}
\caption{Calculated spectra in the B$_{1g}$ and B$_{2g}$
symmetries in the normal and superconducting states (dotted and
full lines respectively). A $d$-wave gap has been assumed:
$\Delta_{\mathbf{k}}=\Delta_0(\text{cos}k_xa-\text{cos}k_ya)$. The
normal state spectra were obtained by setting
$\Delta_{\mathbf{k}}$=0. The corrected spectra are calculated
using the renormalized vertex $\hat{\Gamma}_k$ ($\beta$=0.5 and
$\beta$=1, see text). The frequency dependance of the scattering
rate was taken of marginal Fermi liquid-like form:
$\Sigma''(\omega)=\Sigma''_0+\alpha\omega$ with $\alpha=1$. The
scattering rate was introduced in the Green functions through the
renormalized frequency $\tilde{\omega}=\omega-\Sigma''(\omega)$.}
\label{simul}
\end{figure}

The effect of these fluctuations on the Raman vertex can be
modelled by taking the following phenomenological form for the
renormalized Raman vertex:
$\hat{\Gamma}_{\mathbf{k}}$=$\gamma_{\mathbf{k}}\hat{\tau}_3\text{exp}(-\beta(\text{cos}k_xa-\text{cos}k_ya)^2)$
where $\beta$ is a dimensionless parameter which, in this
framework, depends on the strength of the coupling between the
anti-nodal quasiparticles and the charge suppressing fluctuations.
The calculated responses for both channels,
with and without vertex corrections, in the normal and
superconducting states, are shown in FIG. \ref{simul}. Contrary to
the B$_{2g}$ response, the renormalized B$_{1g}$ response is
strongly suppressed and the 2$\Delta_0$ pair breaking peak becomes
hardly detectable. The theoretical spectra are qualitatively
consistent with the experimental data in underdoped cuprates and
underline the crucial role of vertex corrections on the channel
dependant Raman response.

In summary, we have shown that anisotropic two-particle vertex
renormalizations probed by ERS demonstrate a dichotomy between the
dynamics of nodal and anti-nodal electrons. Our ERS results
augment prior ARPES work to show that manifestations of the
pseudogap hints towards its non-superconducting origin.

We wish to acknowledge helpful discussions with E. Ya. Sherman, Y.
Sidis and P. Bourges, and technical support by A. Forget.

\end{document}